\documentclass[useAMS,usenatbib,a4paper]{mn2e}

\usepackage{psfig}
\usepackage{graphics}

\newcommand\psr{J0218+4232}

\title[PSR \psr\ Distance]{Why the distance of PSR \psr\ does not
  challenge pulsar emission theories}

\author[J.~P.~W.~Verbiest and
D.~R.~Lorimer]{J.~P.~W.~Verbiest$^{1,2}$\thanks{E-mail:
    verbiest@physik.uni-bielefeld.de} and D.~R.~Lorimer$^{3,4}$\\
$^1$Fakult\"at f\"ur Physik, Universit\"at Bielefeld, Postfach 100131,
33501 Bielefeld, Germany\\
$^2$Max-Planck-Institut f\"ur Radioastronomie, Auf dem H\"ugel 69,
53121 Bonn, Germany\\
$^3$Department of Physics \& Astronomy, West Virginia University,
Morgantown, WV 26506, USA\\
$^4$National Radio Astronomy Observatory, Green Bank, WV 24944, USA}

\begin{document}

\date{Accepted. Received; in original form 2014 March xx}

\pagerange{\pageref{firstpage}--\pageref{lastpage}} \pubyear{2014}

\maketitle

\label{firstpage}

\begin{abstract}
  Recent VLBI measurements of the astrometric parameters of the
  millisecond pulsar \psr\ by Du et al.
  have suggested this pulsar is as distant as
  6.3\,kpc. At such a large distance, the large $\gamma$-ray flux
  observed from this pulsar would make it the most luminous
  $\gamma$-ray pulsar known. This luminosity would exceed what can be
  explained by the outer gap and slot-gap pulsar emission models,
  potentially placing important and otherwise elusive constraints on
  the pulsar emission mechanism. We show that the VLBI parallax
  measurement is dominated by the Lutz-Kelker bias. When this bias is
  corrected for, the most likely distance for this pulsar is
  $3.15^{+0.85}_{-0.60}$\,kpc. This revised distance places the luminosity
  of PSR \psr\ into a range where it does not challenge any of the standard
  theories of the pulsar emission mechanism.
\end{abstract}


\section{Introduction}
The millisecond pulsar (MSP) \psr\ was discovered by \citet{nbf+95}
and immediately stood out as a particularly bright MSP with a
relatively high dispersion measure (DM). This DM placed it at a
distance of 5.7\,kpc based on the Galactic electron density model used
at the time \citep{tc93}. Subsequently, \citet{vkb+96} detected the
pulsar in both X-rays and $\gamma$-rays and showed that the
luminosities at these high frequencies were very high, though not
impossibly so. Furthermore, these authors commented that the derived
$\gamma$-ray efficiency was 10\% and could only be predicted by the
outer gap model for pulsar emission, not by the polar cap model. This
situation changed, however, with the advent of a new electron density
model for the Galaxy \citep{cl02}, which dramatically changed the
derived distance down to 2.7\,kpc, in close agreement with the
distance range of 2.5 to 4\,kpc derived from optical observations of
the pulsar's white dwarf companion \citep{bvk03}. At this distance,
the pulsar's luminosity in $\gamma$-rays is still high, but no longer
extreme. As noted by \citet{kbm+03}, this large variability in
distances derived from electron density models, implies that care must
be taken in deriving consequences from $\gamma$-ray luminosities based
on such distances, in particular for PSR~\psr.

The debate on the distance (and thereby $\gamma$-ray luminosity) of
PSR~\psr\ changed again, when \citet{dyc+14} recently presented the
first VLBI measurements of this pulsar's astrometric parameters. They
determined a total proper motion of $6.53\pm 0.08$\,mas/yr and, more
importantly, a parallax of $0.16\pm 0.09$\,mas, translated into a
distance of $6.3^{+8.0}_{-2.3}$\,kpc. 
These results are significant because they make PSR~\psr\ by far the
most luminous $\gamma$-ray pulsar. Furthermore, the VLBI distance and
proper motion imply a transverse speed of $195^{+249}_{-71}$\,km/s,
which is also extreme for the class of MSPs. \citep[derive an average
transverse speed of $87\pm13$\,km/s for this class of pulsar.]{hllk05}

In this paper, we demonstrate that the parallax value reported by
\citet{dyc+14} is dominated by the statistical Lutz-Kelker bias. We
demonstrate that correction for this bias provides a distance that is
more in line with the distance derived from the white-dwarf cooling
models of \citet{bvk03} and with expected pulsar $\gamma$-ray fluxes
and transverse velocities. Our analysis is given in
Section~\ref{sec:LKB}, the implications for the pulsar velocity and
$\gamma$-ray luminosity are detailed in Section~\ref{sec:consequences}
and our conclusions are summarised in Section~\ref{sec:conc}.

\section{Correcting Biases in Parallax Measurements}\label{sec:LKB}
The most important statistical bias that affects pulsar parallax
measurements with limited measurement precision, is the Lutz-Kelker
bias \citep{lk73}. Because of the non-linear scaling of volume with
parallax, the error volume on the lower-parallax end is larger than on
the higher parallax end, resulting in anomalously large parallax
measurements. This effect was investigated by \citet{vlm10}, who found
that parallax measurements with significance below 2-$\sigma$ are
likely dominated by this bias.

\begin{figure}
  \psfig{angle=0.0,width=8.5cm,figure=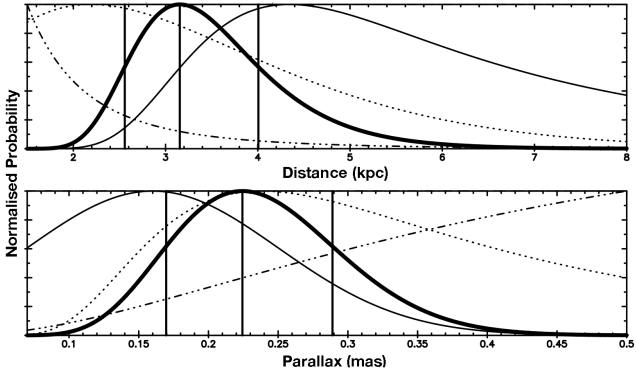}
  \caption{The probability distributions for the parallax and distance
    of PSR~\psr. The combined probability distribution is indicated by
    the thick line; the probability distribution derived from the VLBI
    parallax measurement is given by the thin, continuous line; the
    dotted line shows the probability distribution derived from the
    Galactic distribution of pulsars (the so-called ``volumetric''
    probability); and the triple-dot-dashed line shows the probability
    distribution derived from the pulsar's published flux at
    1.4\,GHz. The most likely value and its 1-$\sigma$ uncertainties
    are indicated by the vertical lines.}
  \label{fig:LKB}
\end{figure}

Applying the bias-correction code of \citet{vwc+12}, we find that the
1.8-$\sigma$ measurement of \citet{dyc+14} is indeed dominated by the
Lutz-Kelker bias. The bias-corrected parallax value (see
Figure~\ref{fig:LKB}) is $0.22^{+0.07}_{-0.05}$\,mas, corresponding to
a distance of $3.15^{+0.85}_{-0.60}$\,kpc\footnote{Note that due
  to the non-linear character of the parallax-to-distance conversion,
  the most likely distance estimate is not the inverse of the most
  likely parallax value, unless in the absence of measurement
  uncertainties.} 

\begin{figure}
  \psfig{angle=0.0,width=8.5cm,figure=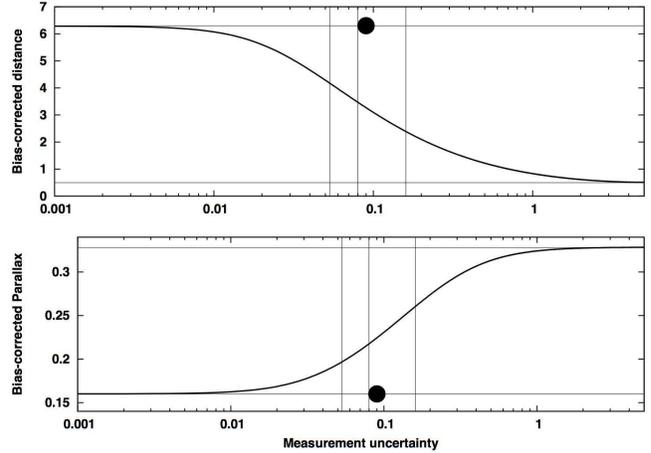}
  \caption{The strength of the bias as a function of the measurement
    uncertainty. For a parallax measurement of 0.16\,mas, the thick
    line shows the bias-corrected distance (top) or parallax (bottom)
    as a function of the measurement uncertainty. The two horizontal
    lines are at the distance and parallax values proposed by
    \citet{dyc+14} and as derived from our analysis.
    The vertical lines indicate 1, 2 and 3-$\sigma$ measurements (from
    right to left) and the recently published VLBI parallax
    measurement of \citet{dyc+14} is shown by the black dot.}
  \label{fig:Evol}
\end{figure}

With improved measurement precision, the impact of these biases will
decrease, as shown in Figure~\ref{fig:Evol}. This indicates that, in
order to reduce the effect of bias to within 0.5\,$\sigma$, a parallax
precision of better than 0.3\,mas would be needed, requiring a factor
three improvement in the published parallax value. This effort would
be worthwhile, though our analysis suggests \citep[as demonstrated
in][]{vlm10}, that it would be more likely that the measured parallax
value would gradually increase as the measurement precision improved.

Figure~\ref{fig:Evol} also demonstrates how this bias would worsen in
case the \citet{dyc+14} parallax uncertainty is underestimated. For
example, if their measurement precision was underestimated even by
only 33\%, the bias-corrected distance would become as low as
2.7\,kpc. 

\section{Consequences of the Bias-Corrected Parallax Value}\label{sec:consequences}
\citet{dyc+14} derive two significant results from their astrometric
measurements. Firstly, they combine their proper motion and parallax
measurements to derive a transverse speed for the pulsar of
$195^{+249}_{-71}$\,km/s. This value is well above the mean MSP
transverse speed
of $87\pm13$\,km/s \citep{hllk05}. Redetermining the transverse speed
of PSR~\psr\ with our bias-corrected parallax value, we obtain
98\,km/s, which lies well within the scatter inherent to the
population.

More importantly, the VLBI distance was used to determine the
$\gamma$-ray luminosity of PSR~\psr. This resulted in $L_{\gamma} =
2.2\times 10^{35}$\,erg/s, which places it more than a factor of three
above the next most luminous $\gamma$-ray MSP, implying a $\gamma$-ray
efficiency in excess of $90\%$. Our bias-corrected distance of
$3.15$\,kpc, however, results in a $\gamma$-ray luminosity of
$5.4\times 10^{34}$\,erg/s, which lies well inside the range of
$\gamma$-ray luminosities for MSPs and a $\gamma$-ray efficiency of
23\%, also comparable to the population at large \citep{aaa+13}.

\section{Conclusions}\label{sec:conc}
We have quantified and corrected biases present in the recent VLBI
parallax to PSR~\psr. Our results are collated in Table~\ref{tab:res}
and are as follows. For the parallax we obtain a most-likely value of
$0.22^{+0.07}_{-0.05}$\,mas and for the distance
$3.15^{+0.85}_{-0.60}$\,kpc. 
We have demonstrated that these bias-corrected values
result in a transverse velocity of 98\,km/s, which is comparable to
values obtained for the rest of the MSP population; and in a
$\gamma$-ray luminosity of $5.4\times 10^{34}$\,erg/s, also in line
with values obtained for other MSPs. The $\gamma$-ray emission of this
pulsar, therefore, does not challenge the outer gap or slot-gap models
for pulsar emission, as claimed by \cite{dyc+14}.

\begin{table}
  \centering
    \caption{Summary of results. Shown are the parallax and distance 
      from the \citet{dyc+14} VLBI measurements (1), from our standard
      bias-correction code described in \citet{vwc+12} (2) and from the
      volumetric and luminosity information alone (i.e. the prior
      information excluding the parallax measurement; 3). 
    }
    \label{tab:res}
    \begin{tabular}{llll}
      \hline
      Method & Parallax (mas) & Distance (kpc) \\ 
      \hline
      (1) VLBI & $0.16\pm 0.09$ & $6.3^{+8.0}_{-2.3}$ \\ 
      (2) Bias-corrected & $0.22^{+0.07}_{-0.05}$ &
      $3.15^{+0.85}_{-0.60}$ \\ 
      (3) vol. \& lum. priors & $0.3^{+0.3}_{-0.1}$ &
      $0.5^{+0.8}_{-0.3}$ \\ 
      \hline
    \end{tabular}
\end{table}

\section*{Acknowledgments}

The authors wish to thank Ben W.\ Stappers and Adam Deller for
discussion of the PSR~J0218+4232 astrometry results, Michael Kramer
for comments on the initial draft and the anonymous referee for a
thorough and critical review.

\bibliographystyle{mn2e}
\bibliography{journals,psrrefs,modrefs,crossrefs}

\label{lastpage}

\end{document}